%% file: Paper.tex
\documentclass[11pt,a4paper]{article}
\usepackage{jheppub}
\usepackage{atlasphysics} 
\usepackage{upgreek}
\input{keys}

\usepackage{preprintcover} 
\PreprintCoverPaperTitle{Time-dependent angular analysis of the decay \Bstbold\  and extraction of \DGs\ and the $CP$-violating weak phase \phis\  by ATLAS}  
\PreprintIdNumber{CERN-PH-EP-2012-182}  
\PreprintCoverAbstract{A measurement of \Bst\ decay parameters, including the  $CP$-violating weak phase \phis\ and  the decay width difference \DGs~is reported,  using \ilumi\  of integrated luminosity collected in 2011 by the ATLAS detector from LHC $pp$ collisions at a centre-of-mass energy  $\sqrt{s} = 7 \TeV$. The mean decay width \Gs\ and the transversity amplitudes $|A_{0}(0)|^2$ and $|A_{\parallel}(0)|^2$ are also measured.  
The values reported for these parameters are:
\begin{alignat*}{3}
\phis & = & \pSfit \,~  & \ \pm & \pSstat \,~~ \rm{(stat.)} & \pm  \ \pSsyst \,~~ \rm{(syst.)}~rad  \nonumber \\
\DGs &  = & ~ \DGfit  & \ \pm  & ~ \DGstat~\rm{(stat.)}  & \pm   ~  \DGsyst~\rm{(syst.)~ps}^{-1} \nonumber \\
\Gs & = & \GSfit & \ \pm & \GSstat ~\rm{(stat.)} & \pm \ \GSsyst ~\rm{(syst.)~ps}^{-1} \nonumber \\
|A_{0}(0)|^2 & = & \Azesqfit & \ \pm & \Azesqstat ~\rm{(stat.)}& \pm \ \Azesqsyst~\rm{(syst.)} \nonumber \\
|A_{\parallel}(0)|^2 & = & \Apasqfit & \ \pm & \Apasqstat ~\rm{(stat.)}&  \pm \ \Apasqsyst~\rm{(syst.)} \nonumber \\
\end{alignat*}
where the values quoted for \phis\ and \DGs\ correspond to the solution compatible with the external measurements to which the strong phase $\delta_{\perp}$  is constrained  and where \DGs\ is constrained to be positive.
 The fraction of $S$-wave $KK$ or $f_{0}$ contamination through the decays \BstoKK\ is measured as well and is found to be consistent with zero. Results for \phis\ and \DGs\ are also presented as 68\%,  90\% and 95\% likelihood contours, which show agreement with Standard Model expectations.  }  
\PreprintJournalName{JHEP}

\title{Time-dependent angular analysis of the decay \Bstbold\  and extraction of \DGs\ and the $CP$-violating weak phase \phis\  by ATLAS}
\author{The ATLAS Collaboration}

\abstract{
 A measurement of \Bst\ decay parameters, including the  $CP$-violating weak phase \phis\ and  the decay width difference \DGs~is reported,  using \ilumi\  of integrated luminosity collected in 2011 by the ATLAS detector from LHC $pp$ collisions at a centre-of-mass energy  $\sqrt{s} = 7 \TeV$. The mean decay width \Gs\ and the transversity amplitudes $|A_{0}(0)|^2$ and $|A_{\parallel}(0)|^2$ are also measured.  
The values reported for these parameters are:
\begin{alignat*}{3}
\phis & = & \pSfit \,~  & \ \pm & \pSstat \,~~ \rm{(stat.)} & \pm  \ \pSsyst \,~~ \rm{(syst.)}~rad  \nonumber \\
\DGs &  = & ~ \DGfit  & \ \pm  & ~ \DGstat~\rm{(stat.)}  & \pm   ~  \DGsyst~\rm{(syst.)~ps}^{-1} \nonumber \\
\Gs & = & \GSfit & \ \pm & \GSstat ~\rm{(stat.)} & \pm \ \GSsyst ~\rm{(syst.)~ps}^{-1} \nonumber \\
|A_{0}(0)|^2 & = & \Azesqfit & \ \pm & \Azesqstat ~\rm{(stat.)}& \pm \ \Azesqsyst~\rm{(syst.)} \nonumber \\
|A_{\parallel}(0)|^2 & = & \Apasqfit & \ \pm & \Apasqstat ~\rm{(stat.)}&  \pm \ \Apasqsyst~\rm{(syst.)} \nonumber \\
\end{alignat*}
where the values quoted for \phis\ and \DGs\ correspond to the solution compatible with the external measurements to which the strong phase $\delta_{\perp}$  is constrained  and where \DGs\ is constrained to be positive.
 The fraction of $S$-wave $KK$ or $f_{0}$ contamination through the decays \BstoKK\ is measured as well and is found to be consistent with zero. Results for \phis\ and \DGs\ are also presented as 68\%,  90\% and 95\% likelihood contours, which show agreement with Standard Model expectations.  
}

\begin{document}
\maketitle

\section{Introduction}
New phenomena beyond the predictions of the Standard Model (SM) may alter $CP$ violation in $B$-decays. A channel that  is expected to be sensitive to  new physics contributions is the decay \Bst. $CP$ violation in the \Bst\ decay occurs due to interference between direct decays and decays  occurring through \BsaBs mixing.  The oscillation frequency of \Bs\ meson mixing is characterized by the mass difference \dms\ of the heavy (\BH) and light (\BL)  mass eigenstates. The $CP$-violating phase \phis\ is defined as the weak phase difference between the \BsaBs mixing amplitude and the $b \rightarrow c \overline{c} s$ decay amplitude.  In the absence of $CP$ violation, the \BH\ state would correspond exactly to the $CP$-odd state and the \BL\ to the $CP$-even state.  In the SM the phase \phis\ is small and can be related to CKM quark mixing matrix elements via the relation $\phis \simeq -2 \beta_{s}$, with $\beta_{s} = \mathrm{arg} [- (V_{ts} V^{*}_{tb})/(V_{cs} V_{cb}^{*}) ]$; a value of  $\phis \simeq -2 \beta_{s} = -0.0368 \pm 0.0018 $ rad \cite{PHISM} is predicted in the SM. Many new physics models predict large  \phis\ values whilst satisfying all existing constraints, including the precisely measured value of  \dms\  \cite{dmsTevatron,LHCboscil}.

Another physical quantity involved in \BsaBs mixing is the width difference $\DGs = \GL - \GH$ of \BL\ and \BH.  Physics beyond the SM is not expected to affect \DGs\ as significantly as \phis~\cite{Nierste}. Extracting \DGs\ from data is nevertheless useful as it allows theoretical predictions to be tested~\cite{Nierste}.

The decay of the pseudoscalar \Bs\ to the vector-vector  final-state $J/\psi\phi$  results in an admixture of $CP$-odd and $CP$-even states, with orbital angular momentum $L = 0$, 1 or 2. The final states with orbital angular momentum $L = 0$ or 2 are $CP$-even while the state with $L = 1$ is $CP$-odd.  No flavour tagging to distinguish between   the initial \Bs\ and \aBsn\ states is used in this analysis; the $CP$ states are separated statistically through the time-dependence of the decay and angular correlations amongst the final-state particles. 

In this paper,  measurements of \phis, the average decay width $\Gs = (\Gamma_L + \Gamma_H) / 2$ and the  value  of \DGs, using the fully reconstructed decay \Bsto \ are presented. Previous measurements of these quantities have been reported by the CDF and D\O\ collaborations \cite{CDF:2011af, Abazov:2011ry}  and recently by the LHCb collaboration \cite{LHCb:2011aa}. The analysis presented here uses data collected by the ATLAS detector from LHC $pp$ collisions running at $\sqrt{s} = 7 \TeV$ in 2011, corresponding to an integrated luminosity of approximately $4.9~ \ifb$. 

\section{ATLAS detector and Monte Carlo simulation} \label{sec:sample}
The ATLAS experiment \cite{ATLAS} is a multipurpose particle physics detector with a forward-backward symmetric cylindrical geometry and near $4\pi$ coverage in solid angle. The inner tracking detector (ID) consists of a silicon pixel detector, a silicon microstrip detector  and a transition radiation tracker. The ID is surrounded by a thin superconducting solenoid providing a 2\,T axial magnetic field, and by high-granularity liquid-argon (LAr) sampling electromagnetic calorimeter. An iron-scintillator tile calorimeter provides hadronic coverage in the central rapidity range. The end-cap and forward regions are instrumented with LAr calorimeters for both electromagnetic and hadronic measurements. The muon spectrometer (MS) surrounds the calorimeters and consists of three large superconducting toroids with eight coils each, a system of tracking chambers, and detectors for triggering.

The muon and tracking systems are of particular importance in the reconstruction of \B\ meson candidates. Only data where both systems were operating correctly and where the LHC beams were declared to be stable are used. The data were collected during a period of rising instantaneous luminosity at the LHC, and the trigger conditions varied over this time. 

The triggers used to select events for this analysis are based on identification of a \Jpsitomm decay, with either a $4\;\mathrm{GeV}$ transverse momentum\footnote{The ATLAS coordinate system and the definition of transverse momentum are described in reference~\cite{ATLAS}} (\pT) threshold for each muon or an asymmetric configuration that applies a higher \pT\ threshold ($4-10\GeV$) to one of the muons and a looser muon-identification requirement (\pT\ threshold below $4\;\mathrm{GeV}$) to the second one.

Monte Carlo (MC) simulation is used to study the detector response, estimate backgrounds and model systematic effects. For this study, 12 million MC-simulated \Bst\ events were generated using PYTHIA \cite{Pythia} tuned with~recent ATLAS data~\cite{ATL-PHYS-PUB-2011-009}. No \pT\ cuts were applied at the generator level.  Detector responses for these events were simulated using an ATLAS simulation package based on GEANT4~\cite{Atlfast,Geant4}.  In order to take into account the varying trigger configurations during data-taking, the MC events were weighted to have the same trigger composition as the collected collision data.  Additional samples of the  background decay $\Bd \ra J/\psi K^{0*}$ as well as the more general $bb \ra J/\psi X$ and $pp \ra J/\psi X$ backgrounds were also simulated using PYTHIA.

\section{Reconstruction and candidate selection}
Events passing the trigger and the data quality selections described in section~\ref{sec:sample} are required to pass the following additional criteria: the event must contain at least one reconstructed primary vertex built from at least four  ID tracks in order to be considered in the subsequent analysis; the event must contain at least one pair of oppositely charged muon candidates that are reconstructed using two algorithms that combine the information from the MS and the ID \cite{Aad20119}.   
In this analysis the muon track parameters are taken from the ID measurement alone, since the precision of  the measured track parameters for muons in the  \pT\ range of interest for this analysis is dominated by the ID track reconstruction.  The pairs of muon tracks are refitted to a common vertex and accepted for further consideration if the fit results in \cutChiJpsi. The invariant mass of the muon pair is calculated from the refitted track parameters.  
To account for  varying  mass resolution, the \Jpsi\ candidates are divided into three subsets  according to the pseudorapidity $\eta$ of the muons. A maximum likelihood fit is used to extract the $J/\psi$ mass and the corresponding resolution for these three subsets.  When both muons have $|\eta| <  1.05$, the di-muon invariant mass must fall in the range \cutMassJpsiBB \ to be accepted as a  \Jpsi\ candidate. When one muon has $1.05  < |\eta| <  2.5 $ and the other muon  $|\eta| <  1.05$, the corresponding signal region is \cutMassJpsiEB. For the third subset, where both muons have $1.05  < |\eta| <  2.5 $, the signal region is \cutMassJpsiEE .  In each case the signal region is defined so as to  retain 99.8\% of the  \Jpsi\ candidates identified in the fits.  

The candidates for  \phiKK are reconstructed from all pairs of oppositely charged tracks with $\pT > 0.5$~\GeV \ and $|\eta| < 2.5$ that are not identified as muons. Candidates  for  \Bsto\ are sought by fitting the tracks for each combination of \Jpsitomm\ and \phiKK to a common vertex.  All four tracks are required to have at least one hit in the pixel detector and at least four hits in the silicon strip detector. The fit is further constrained by fixing the invariant mass calculated from the two muon tracks to the world average \Jpsipa mass~\cite{PDG}. These quadruplets of tracks are accepted for further analysis if the vertex fit has a  \cutChiBs, the fitted  \pT\ of each track from \phiKK\ is greater than 1 GeV and the invariant mass of the track pairs (under the assumption that they are kaons) falls within the interval \cutMassPhi.
In total $131\mathrm{k}$ \Bs\ candidates are collected within a mass range of $5.15 < m(\Bs) < 5.65 \GeV$ used in the fit.

For each \Bs\ meson candidate the proper decay time $t$ is determined by the expression:
\begin{equation*}
  t = \frac{ L_{xy}\ M_{{}_{B}} }     {c \ p_{ \mathrm{T}_{B}  }}, 
\end{equation*}
where $p_{ \mathrm{T}_{B}  }$ is the  reconstructed transverse momentum of the \Bs\ meson candidate and  $M_{{}_{B}}$ denotes  the world average mass value \cite{PDG} of the \Bs\ meson (\BspdgNoerr). The transverse decay length 
$L_{xy}$ is the displacement in the transverse plane of the  \Bs\ meson  decay vertex with respect to the primary vertex, projected onto the direction of \Bs\ transverse momentum.
The position of the primary vertex used to calculate this quantity is refitted following the removal of the tracks used to reconstruct the \Bs\ meson candidate.

For the selected events the average number of pileup interactions is 5.6, necessitating  a choice of the best candidate for the primary vertex at which the \Bs\  meson is  produced.  The variable used  is a three-dimensional impact parameter $d_0$, which is  calculated as the distance between the line extrapolated from the reconstructed \Bs\ meson vertex in the~direction of the \Bs\ momentum, and  each primary vertex candidate. The chosen primary vertex is the one with the smallest $d_0$. 
  Using MC simulation it is shown that the fraction of \Bs \ candidates which are assigned the wrong primary vertex is less than 1\% and that the corresponding effect on the final results is negligible. No \Bs\ meson lifetime cut is applied in the analysis. 

\section{Maximum likelihood fit}
An unbinned maximum likelihood fit is performed on the selected events to extract the parameters of the \Bsto\ decay.  The fit uses information about the reconstructed mass $m$, the measured proper decay time $t$, the measured mass and proper decay time uncertainties $\sigma_m$ and $\sigma_t$,  and the transversity angles $\Omega$ of each \Bst\ decay candidate.  There are three transversity angles; $\Omega = (\theta_T,\psi_T,\varphi_T)$ and these are defined in section \ref{sec:sigPDF}.

The likelihood function is defined as a combination of the signal and background probability density functions as follows:
\begin{alignat}{3} \label{eq:likelihood}
\ln~{\cal L}  =  \sum_{i=1}^N \{  w_i \cdot \mathrm{ln} (  f_{\textrm{s}} \cdot   {\cal F}_{\textrm{s}}(  m_{i}, t_{i}, \Omega_{i} )
 +  f_{\textrm{s}} \cdot f_{B^0} \cdot {\cal F}_{B^0}  (  m_{i}, t_{i}, \Omega_{i} ) \nonumber \\
 + ( 1 -  f_{\textrm{s}} \cdot (1+ f_{B^0 } )    )  {\cal F}_{\textrm{bkg}} (  m_{i}, t_{i}, \Omega_{i} )  ) \} 
+  \mathrm{ln} P({\delta_\perp})  \nonumber \\
\end{alignat}
where $N$ is the number of selected candidates, $w_i$ is a weighting factor to account for the trigger efficiency (described in section~\ref{sec:Teff}), $f_\textrm{s}$ is the fraction of signal candidates,  $f_{B^0}$ is the fraction of  peaking $B^0$  meson background events (described in section \ref{sec:Bd}) calculated relative to the number of signal events; this parameter is  fixed  in the likelihood fit. The  mass $m_{i}$, the proper decay time $t_{i}$ and the decay angles $\Omega_{i} $ are the values measured from the data for each event $i$.  ${ \cal F}_{\textrm{s} }$, ${\cal F}_{B^0}$ and  ${\cal F}_{\textrm{bkg} }$  are the probability density functions (PDF) modelling the signal, the specific $B^0$  background and  the  other background distributions, respectively.  $P({\delta_\perp})$ is a constraint on the strong phase $\delta_\perp$.  A detailed description of the PDF functions and other terms in the equation \ref{eq:likelihood} is given in sections \ref{sec:sigPDF}~$-$~\ref{sec:Teff}. 

\subsection{Signal PDF\label{sec:sigPDF}} 
The  PDF describing  the signal events, ${\cal F}_{\textrm{s}}$,   has the form of a product of PDFs for each quantity measured from the data:
\begin{eqnarray} \label{eq:likelihoodSig}
  {\cal F}_{\textrm{s}}(  \it{m_{i}, t_{i}, \Omega_{i}} ) &=&  P_\textrm{s}(m_{i}|\sigma_{m_{i}}) \cdot P_\textrm{s}(\sigma_{m_{i}})\cdot P_\textrm{s}(\Omega_{i}, t_{i}|\sigma_{t_{i}}) \cdot P_\textrm{s}(\sigma_{t_{i}}) \cdot A(\Omega_i,{\it p_{\mathrm{T}i}})  \cdot P_\textrm{s}({\it p_{\mathrm{T}i}}) \nonumber \\
 \end{eqnarray}
 The terms $P_\textrm{s}(m_{i}|\sigma_{m_{i}})$, $P_\textrm{s}(\Omega_{i}, t_{i}|\sigma_{t_{i}})$ and  $A(\Omega_i,{\it p_{\mathrm{T}i}})$ are 
 explained in the current section,  and the remaining per-candidate uncertainty terms $P_\textrm{s}(\sigma_{m_{i}})$, $P_\textrm{s}(\sigma_{t_{i}})$ and $P_\textrm{s}({\it p_{\mathrm{T}i}})$  are described in section \ref{sec:Punzi}.
Ignoring detector effects, the joint distribution for the decay time $t$ and the transversity angles $\Omega$ for the \Bsto\ decay is given by the differential decay rate \cite{CDFAngles}:
\begin{equation}\label{eq:ang2}
\frac{d^4 \Gamma}{dt\ d\Omega}= \sum_{k=1}^{10} {\cal O}^{(k)}(t) g^{(k)}(\theta_T,\psi_T,\varphi_T), 
\end{equation}
where  ${\cal O}^{(k)}(t)$ are the time-dependent amplitudes and $g^{(k)}(\theta_T,\psi_T,\varphi_T)$ are the angular functions,  given in table \ref{tab:SigPDF}.  The time-dependent amplitudes are slightly different for decays of mesons that were initially \aBsn.  As an untagged analysis is performed here, all \Bs\ meson candidates are assumed to have had an equal chance of initially being either a particle or anti-particle.  This leads to a significant simplification of the time-dependent amplitudes as any terms involving the mass splitting $\Delta m_s$ cancel out.  These simplified time-dependent amplitudes are given in table \ref{tab:SigPDF}.  $A_\perp(t)$ describes a $CP$-odd final-state configuration while both $A_0(t)$ and $A_\parallel(t)$ correspond to $CP$-even final-state configurations.   $A_S$ describes the contribution of $CP$-odd $B_s\rightarrow J/\psi K^+K^- (f_0)$, where the non-resonant $KK$ or $f_0$ meson is an $S$-wave state.  The corresponding amplitudes are given in the last four lines of  table \ref{tab:SigPDF}  ($k$=7-10) and follow the convention used  in previous analysis~\cite{LHCb:2011aa}. The likelihood is independent of the invariant $KK$ mass distribution. 

The equations are normalised such that the squares of the amplitudes sum to unity; three of the four amplitudes are fit parameters and  $|A_\perp(0)|^2$ is determined according to this constraint.

The angles ($\theta_T,\psi_T,\varphi_T$), are defined in the rest frames of the final-state particles.  The $x$-axis is determined by the direction of the $\phi$ meson in the $J/\psi$ rest frame, the $K^{+}K^{-}$  system defines the $xy$ plane, where $p_{y}(K^{+})>0$. The three angles are defined:
\begin{itemize}
\item$\theta$, the angle between $p(\mu^{+})$ and the $xy$ plane, in the $J/\psi$ meson rest frame
\item $\varphi$, the angle between the $x$-axis and $p_{xy}(\mu^{+})$, the projection of the $\mu^+$ momentum in the $xy$ plane, in the $J/\psi$ meson rest frame
\item $\psi$, the angle between $p(K^{+})$ and $-p(J/\psi)$ in the $\phi$ meson rest frame 
\end{itemize}

\begin{table}[htdp]
\small
\begin{center}
\begin{tabular}{|c| l | l |}
\hline
$k$ & ${\cal O}^{(k)}(t)$ & $g^{(k)}(\theta_T,\psi_T,\varphi_T)$ \\
\hline \hline
1 & $\frac{1}{2} |A_0(0)|^2 \left[\left(1+\cos\phis\right) e^{-\Gamma_{\rm L}^{(s)} t} + \left(1-\cos\phis\right) e^{-\Gamma_{\rm H}^{(s)} t} \right]$ & $2 \cos^2\psi_T(1 - \sin^2\theta_T\cos^2\varphi_T)$ \\
2 & $\frac{1}{2}|A_{\|}(0)|^2 \left[\left(1+\cos\phis\right) e^{-\Gamma_{\rm L}^{(s)} t} + \left(1-\cos\phis\right) e^{-\Gamma_{\rm H}^{(s)} t} \right]$ & $\sin^2\psi_T(1 - \sin^2\theta_T\sin^2\varphi_T)$ \\
3 & $\frac{1}{2} |A_{\perp}(0)|^2 \left[\left(1- \cos\phis\right)e^{-\Gamma_{\rm L}^{(s)} t} + \left(1+\cos\phis\right)e^{-\Gamma_{\rm H}^{(s)} t} \right]$ & $\sin^2\psi_T\sin^2\theta_T$ \\
4 & $\frac{1}{2}|A_0(0)||A_{\|}(0)|\cos\delta_{||} $ & $\frac{1}{\sqrt{2}}\sin2\psi_T\sin^2\theta_T\sin2\varphi_T $ \\
 & \hfill $\left[\left(1+\cos\phis\right)e^{-\Gamma_{\rm L}^{(s)} t} + \left(1-\cos\phis\right)e^{-\Gamma_{\rm H}^{(s)} t} \right]$&  \\
5 & $ \frac{1}{2}|A_{\|}(0)||A_{\perp}(0)| \left(e^{-\Gamma_{\rm H}^{(s)} t}-e^{-\Gamma_{\rm L}^{(s)} t}\right)\cos(\delta_{\perp} - \delta_{||})\sin\phis$ & $\sin^2\psi_T\sin2\theta_T\sin\varphi_T$ \\
6 & $- \frac{1}{2}|A_{0}(0)||A_{\perp}(0)|\left(e^{-\Gamma_{\rm H}^{(s)} t}-e^{-\Gamma_{\rm L}^{(s)} t} \right)\cos\delta_{\perp}\sin\phis$ & $\frac{1}{\sqrt{2}} \sin2\psi_T\sin2\theta_T\cos\varphi_T$ \\
7 &$\frac{1}{2} |A_{S}(0)|^2 \left[\left(1- \cos\phis\right)e^{-\Gamma_{\rm L}^{(s)} t} + \left(1+\cos\phis\right)e^{-\Gamma_{\rm H}^{(s)} t} \right]$ &$ \frac{2}{3}\left(1-\sin^2\theta_T\cos^2\varphi_T\right) $ \\
8 &$- \frac{1}{2}|A_{S}(0)||A_{\parallel}(0)|\left(e^{-\Gamma_{\rm H}^{(s)} t}-e^{-\Gamma_{\rm L}^{(s)} t} \right)\sin(\delta_{\parallel}-\delta_{S})\sin\phis$ & $\frac{1}{3} \sqrt{6} \sin\psi_T\sin^2\theta_T\sin 2\varphi_T$ \\
9 &$\frac{1}{2}|A_{S}(0)| |A_{\perp}(0)| $ & $\frac{1}{3} \sqrt{6} \sin\psi_T\sin 2\theta_T\cos\varphi_T$ \\
 &\hfill $\left[\left(1- \cos\phis\right)e^{-\Gamma_{\rm L}^{(s)} t} + \left(1+\cos\phis\right)e^{-\Gamma_{\rm H}^{(s)} t} \right]\sin(\delta_{\perp}-\delta_{S})$ &  \\
10 &$- \frac{1}{2}|A_{0}(0)||A_{S}(0)|\sin(-\delta_{S})\left(e^{-\Gamma_{\rm H}^{(s)} t}-e^{-\Gamma_{\rm L}^{(s)} t} \right)\sin\phis$ &$\frac{4}{3}\sqrt{3}\cos\psi_T\left(1-\sin^2\theta_T\cos^2\varphi_T\right)$ \\
\hline
\end{tabular}
\end{center}
\caption{Table showing the ten time-dependent amplitudes, ${\cal O}^{(k)}(t)$ and the functions of the transversity angles $g^{(k)}(\theta_T,\psi_T,\varphi_T)$. The amplitudes $|A_0(0)|^2$ and $|A_{\parallel}(0)|^2$ are for the $CP$-even components of the \Bst\ decay.   $|A(0)_{\perp}|^2$ is the $CP$-odd amplitude. They have corresponding strong phases $\delta_{0}$, $\delta_{\parallel}$ and $\delta_{\perp}$; by convention $\delta_0$ is set to be zero.  The $S$-wave amplitude $|A_S(0)|^2$ gives the fraction of \BstoKK\ and has a related strong phase $\delta_S$. \label{tab:SigPDF}}
\end{table}

It can be seen from table~\ref{tab:SigPDF}, that  in the untagged analysis used in this study the time-dependent amplitudes depending on $\delta_{\perp}$ (${\cal O}^{(k)}(t), k=5,6$) are multiplied by $\sin\phis$.
Previous measurement by LHCb ref.~\cite{LHCb:2011aa} showed that $\phis$ is close to zero ($0.15 \pm 0.18 \pm 0.06$)~rad.
For such a small value of $\phis$ the untagged analysis is not sensitive to $\delta_{\perp}$.
A Gaussian constraint to the best measured value, $\delta_{\perp} = (2.95 \pm 0.39)$~rad \cite{LHCb:2011aa}, is therefore applied by adding a Gaussian function term $P({\delta_\perp})$ into the likelihood fit.

The signal PDF, $P_\textrm{s}(\Omega_{i}, t_{i}|\sigma_{t_{i}})$ must  take into account the time resolution  and thus each time-dependent element in  table \ref{tab:SigPDF} is  convoluted with a Gaussian function. This convolution is performed numerically on an event-by-event basis where the width of the Gaussian is the proper decay time uncertainty $\sigma_{t_{i}}$,  multiplied by an overall scale factor to account for any mis-measurements. 

The angular sculpting of the detector and kinematic cuts on the angular distributions is included in the likelihood function through $A(\Omega_i, {\it p_{\mathrm{T}i}})$. This is calculated using a four-dimensional binned acceptance method, applying an event-by-event efficiency according to the transversity angles ($\theta_T,\psi_T,\varphi_T$) and the \pT\ of the \Bs. The acceptance was calculated from the \Bst\ MC events. In the likelihood function, the acceptance is treated as an angular sculpting PDF, which is multiplied by   the time- and angular-dependent PDF describing the \Bsto\ decays.   Consequently, the complete angular function must be normalised as a whole as both the acceptance and the time-angular decay PDFs depend on the transversity angles. This normalisation is performed numerically in the likelihood fit. 

The signal mass PDF, $P_\textrm{s}(m_{i})$, is modelled  as a single Gaussian function smeared with an event-by-event mass resolution $\sigma_{m_{i}}$, see figure~\ref{fig:TimeErr}, which is scaled using a  factor to account for mis-estimation of the mass errors. The PDF is normalised over the range $5.15 < m(\Bs) < 5.65 \GeV$.
\begin{figure}[htbp]
\begin{center}
\includegraphics[width=0.45\textwidth]{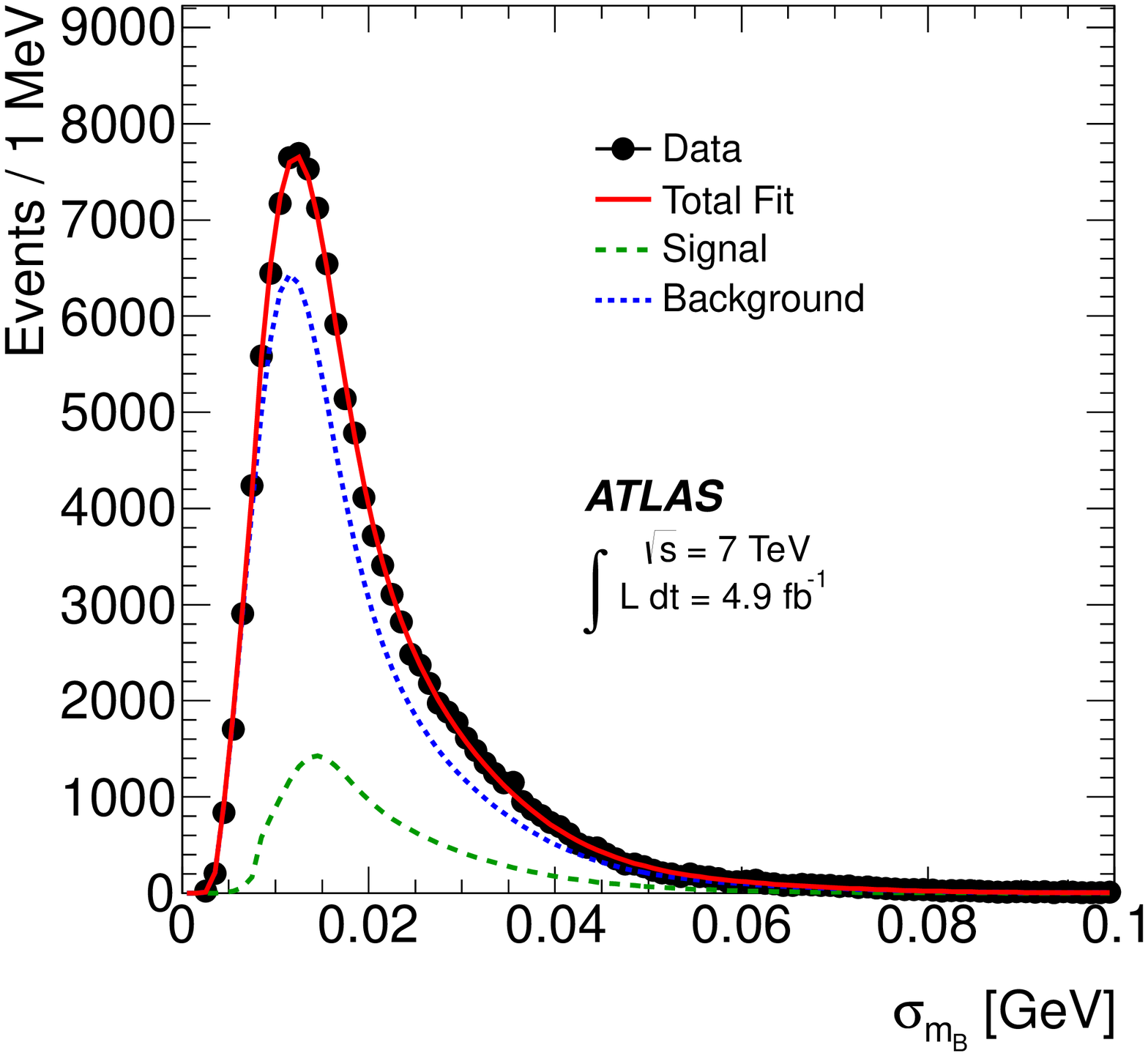}
\includegraphics[width=0.45\textwidth]{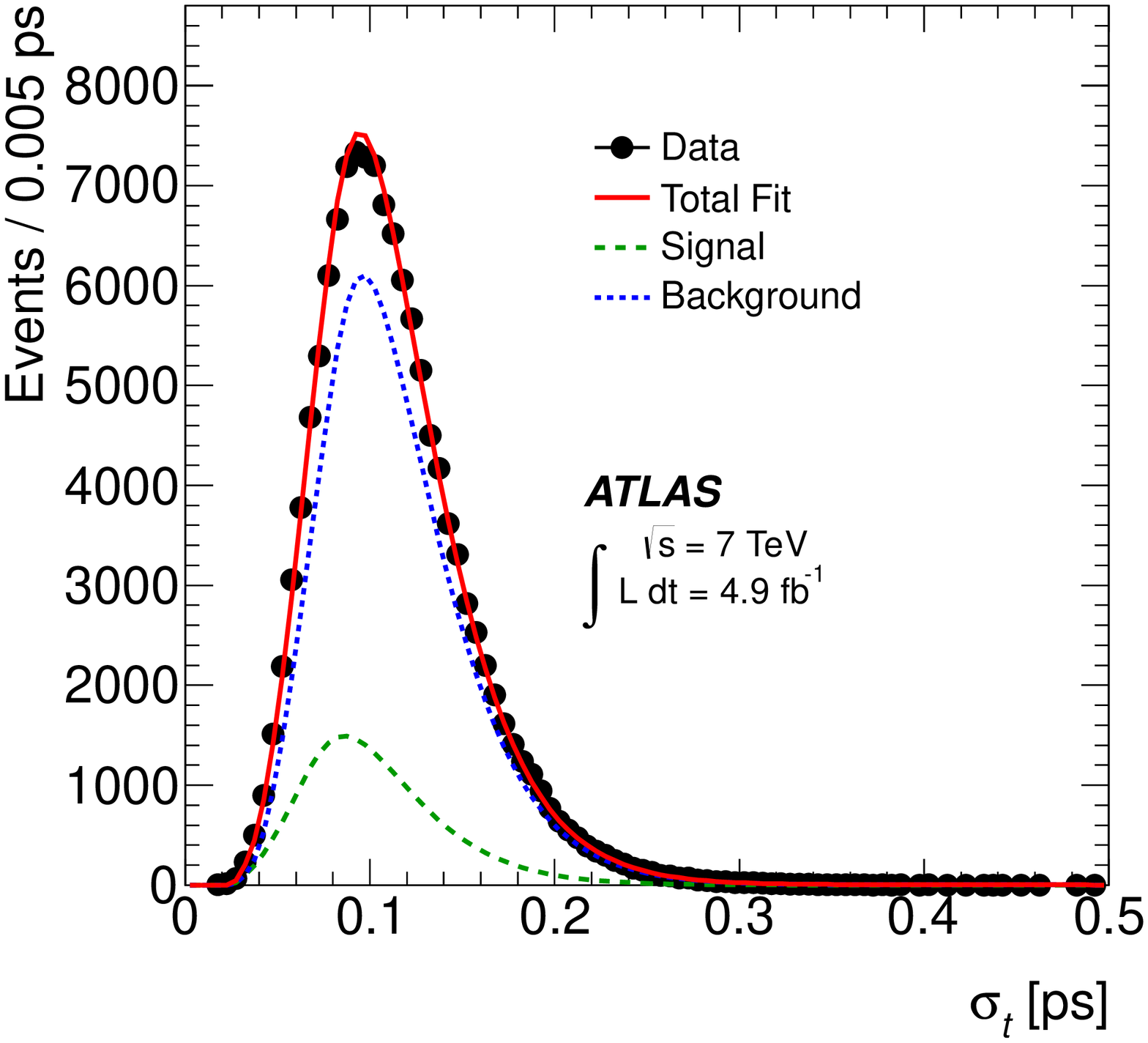}
\caption{Left: mass uncertainty distribution for data,  the fits to the background  and the signal fractions  and the sum of the two fits. Right: proper decay time uncertainty distribution for data,  the fits to the background  and the signal fractions  and the sum of the two fits. }
\label{fig:TimeErr}
\end{center}
\end{figure}

\subsection{Specific \Bd \  background\label{sec:Bd}}
The \Bsto\ sample is contaminated with mis-reconstructed \Bdt\ and $\Bd\rightarrow J/\psi K^{+}\pi^{-}$ (non-resonant) decays, where the final-state pion is mis-identified as a kaon. The two components of the background are referred to as \Bd\ reflections, since the \Bd\ is reconstructed as a \Bs\ meson and therefore lies within the \Bs\ meson mass window rather than in the usual \Bd\ mass range.
The fractions of these components are fixed in the likelihood fit to values $(6.5 \pm 2.4)\%$ and $(4.5 \pm 2.8) \% $
respectively. These values are calculated from the relative production fractions of the \Bs \ and \Bd \  mesons and their decay probabilities taken from the PDG values~\cite{PDG} and from their selection efficiencies, which are determined  from MC events. The  corresponding uncertainties  are dominated by uncertainties in the decay probabilities. 

 Mis-reconstructed \Bd\ decays are treated as part of the background and are described by a dedicated PDF:
\begin{eqnarray} \label{eq:likelihoodBd}
  {\cal F}_{B^0}(  \it{m_{i}, t_{i}, \Omega_{i}} ) & = & P_{B^0}(m_{i})\cdot P_\textrm{s}(\sigma_{m_{i}})\cdot P_{B^0}(t_{i}|\sigma_{t_{i}}) \nonumber \\ 
                                                   & \cdot & P_{B^0 }(\theta_T)\cdot P_{B^0 }(\varphi_T) \cdot P_{B^0 }(\psi_T)  \cdot  P_\textrm{s}(\sigma_{t_{i}}) \cdot P_\textrm{s}({\it p_{\mathrm{T}i}})
\end{eqnarray}

The mass is described by the $P_{B^0}(m_{i})$ term  in the form of  a Landau function due to the distortion caused by the incorrect mass assignment.  The  decay time is described in the term $P_{B^0}(t_{i}|\sigma_{t_{i}})$  by an exponential smeared with event-by-event Gaussian errors.  The transversity angles are described using the same functions as the other backgrounds but with different values for the parameters obtained from the fit to MC data. 
The terms $P_\textrm{s}(\sigma_{m_{i}})$, $P_\textrm{s}(\sigma_{t_{i}})$ and $P_\textrm{s}({\it p_{\mathrm{T}i}})$ are described in section \ref{sec:Punzi}. 
All the~PDFs describing these \Bd\ reflections have fixed shapes determined from~the~MC studies.

\subsection{Background PDF}
The background PDF has the following composition:
\begin{eqnarray} \label{eq:likelihoodBkg}
  {\cal F}_{\textrm{bkg}}(  \it{m_{i}, t_{i}, \Omega_{i}} ) & =  & P_\textrm{b}(m_{i})\cdot P_\textrm{b}(\sigma_{m_{i}}) \cdot P_\textrm{b}(t_{i}|\sigma_{t_{i}}) \nonumber \\ 
                                                            & \cdot & P_\textrm{b}(\theta_T)\cdot P_\textrm{b}(\varphi_T) \cdot P_\textrm{b}(\psi_T)\cdot P_\textrm{b}(\sigma_{t_{i}})  \cdot P_\textrm{b}({\it p_{\mathrm{T}i}})
\end{eqnarray}
The proper decay time function $P_\textrm{b}(t_{i}|\sigma_{t_{i}})$ is parameterised as a prompt peak modelled by a Gaussian distribution, two positive exponentials and a negative exponential. This function is smeared with the same resolution function as the signal decay time-dependence. The prompt peak models the combinatorial background events, which are expected to have reconstructed lifetime distributed around zero. The two positive exponentials represent a  fraction of longer-lived backgrounds with non-prompt $J/\psi$, combined with hadrons from the primary vertex or  from a  $ B/D$  hadron in the same event.  The negative exponential takes into account events with poor vertex resolution.  

The shape of the background angular distributions, $P_\textrm{b}(\theta_T)$, $P_\textrm{b}(\varphi_T)$, and $P_\textrm{b}(\psi_T)$  arise primarily from detector and kinematic sculpting.  These are  described by the following empirically determined functions: 

\begin{eqnarray}
 f(\cos\theta_T) & = & \ \frac{a_{0} - a_{1}\cos^{2}( \theta_T) + a_{2}\cos^{4}(\theta_T)}{2a_{0} - 2a_{1}/3  + 2a_{2}/5  }  \nonumber \\ 
 f(\varphi_T) & = & \ \frac{1 + b_{1}\cos(2\varphi_T+ b_{0})}{2\pi}   \nonumber \\
 f(\cos\psi_T) &  = & \  \frac{c_{0} + c_{1}\cos^{2}(\psi_T)}{2c_{0} + 2c_{1}/3} \nonumber 
\end{eqnarray}
\\
\\
They are initially fitted to data from the \Bs\ mass sidebands only, to find reasonable starting values for $a_{0,1,2}$, $b_{0,1}$ and $c_{0,1}$, then allowed to float freely in the full likelihood fit. The \Bs\ mass sidebands, $(5.150 - 5.317)\GeV$ and $(5.417 - 5.650)\GeV$,
are defined to retain 0.02\%  of signal events identified in the fit. The correlations between the background angular shapes are neglected, but a systematic error arising from this simplification is evaluated in section \ref{sec:Systematics}. The background mass model, $P_\textrm{b}(m)$ is a linear function.

\subsection{Time and mass uncertainties of signal and background\label{sec:Punzi} }
The event-by-event  proper decay time and mass uncertainty distributions differ significantly for signal and background, as shown in figure \ref{fig:TimeErr}. The background PDFs cannot be factorized and it is necessary to include extra PDF terms describing the error distributions in the likelihood function to avoid significant biases~\cite{Punzi}.

The signal and background time and mass error distributions are described with  Gamma functions:
\begin{equation}
P_{\rm s,b}(\sigma_{t(m)_{i}}) = \frac{(\sigma_{t(m)_{i}} - c )^{a_{\rm s,b}}e^{-(\sigma_{t(m)_{i}} - c)/b_{\rm s,b}}}{b_{\rm s,b}^{a_{\rm s,b}+1}\Gamma(a_{\rm s,b}+1)}
\label{eqn:timeErr}
\end{equation}
where $a_{\rm s,b}$ and $b_{\rm s,b}$ are constants fitted from ($b$) sideband and ($s$) sideband-subtracted signal and fixed in the likelihood fit. 
Since $P_{\rm s,b}(\sigma_{t(m)_{i}})$ depend on transverse momentum of the \Bs\ meson, they were determined in six selected \psubt bins, the choice of which is reflecting the natural \psubt dependence of the detector resolution.

The same treatment is used for \Bs \psubt signal and background, by introducing additio-nal terms $P_\textrm{s}({\it p_{\mathrm{T}i}})$ and $P_\textrm{b}({\it p_{\mathrm{T}i}})$ into the PDF.
These are described using the same functions as $P_{\rm s,b}(\sigma_{t(m)_{i}})$ but with different values for the parameters obtained from the fit to sideband and sideband-subtracted signal \psubt distributions.
 
\subsection{Muon trigger time-dependent efficiency\label{sec:Teff}}
It has been observed that the muon trigger biases the transverse impact parameter of muons toward smaller values. The trigger selection efficiency was measured in data and MC simulation using a tag-and-probe method~\cite{ATLAS-CONF-2012-055}. To account for this efficiency in the fit, the events are re-weighted by a factor $w$:
\begin{equation}
  w =  e^{   -|t|/ ( \tau_{\mathrm{sing}}+\epsilon )     }   /     e^{   -|t|/ \tau_{\mathrm{sing}}    }     
   \label{LifeTimeWeight}
\end{equation}
where the $\tau_{\mathrm{sing}}$ is a single \Bs\ lifetime measured before the correction, using unbinned mass-lifetime maximum likelihood fit.  The weight form and the factor  $\epsilon  = 0.013 \pm 0.004\;\mathrm{ps}$ are determined using MC events by comparing  the \Bs\ lifetime distribution of an unbiased sample with the lifetime distribution obtained after including the dependence of the trigger efficiency  on the muon transverse impact parameter as measured  from the data.  The value of $\epsilon$ is determined as the difference of exponential fits to the two distributions.  The uncertainty $0.004\;\mathrm{ps}$,  which reflects the precision of the tag-and-probe method, is used to assign a systematic error due to this time efficiency correction.

\section{Systematic uncertainties \label{sec:Systematics}}
Systematic uncertainties are assigned by considering several effects that are not accounted for in the likelihood fit.  These are described below.  
\begin{itemize}
\item {\bf Inner Detector Alignment:} Residual misalignments of the ID affect the impact parameter distribution with respect to the 
primary vertex. The effect of this residual misalignment on the measurement is estimated using events simulated with perfect and distorted ID
geometries. The distorted geometry is produced by moving detector components to match the observed small shifts in data. The observable
of interest is the impact parameter distribution with respect to the primary vertex as a function of $\eta$ and $\phi$. The mean value of this
impact parameter distribution for a perfectly aligned detector is expected to be  zero and in data a maximum deviation of less than 10 $\upmu$m is observed. The difference between the measurement using simulated events reconstructed with a perfect geometry compared to the distorted geometry is used to assess the systematic uncertainty.

\item {\bf Angular acceptance method:} The angular acceptance is calculated from a binned fit to MC data. In the kinematical region used in this analysis, the angular acceptance varies with the transversity angles by about $\pm$10\%. The statistical error in the acceptance is smaller than 1\% in any bin, and data driven analyses show that systematic uncertainties in modelling Êdetector and reconstruction are also at the level of 1\%  ~\cite{IDPerf, MUPerf}. ÊPossible dependences of the results on the choice of the binning are tested by varying bin widths and central values. Taking all these arguments into consideration, the systematic uncertainties due to detector acceptance are found to be negligible.

\item {\bf Trigger efficiency:}  To correct for the trigger lifetime bias the events are re-weighted according to equation~\ref{LifeTimeWeight}. The uncertainty in the parameter $\epsilon$ is used to estimate the systematic uncertainty due to the time efficiency correction.

\item {\bf Fit model:} Pseudo-experiments are used to estimate systematic uncertainties. In a first test, the results of 1000 pseudo-experiments are Êcompared to the generated values, and the average of the differences are taken as systematic uncertainties. Additional sets of 1000 pseudo-experiments are generated with variations in the signal and background mass model, resolution model, background lifetime and background angles models, as discussed below. These sets are analysed with the default model, and average deviations in the results of the fit are taken as additional systematic errors. The following variations are considered:  
\begin{itemize}
\item The signal mass distribution is generated using a sum of two Gaussian functions. Their relative fractions and widths are  determined  from a likelihood fit to data.  In the PDF for this fit, the mass of each event  is  modelled by two different Gaussians with widths  equal to  products of the scale factors  multiplied by a per-candidate mass error. 
\item The background mass is generated from an exponential function. The default fit uses a linear model for the mass of background events.
\item Two different scale factors instead of one are used to generate the lifetime uncertainty.
\item The values used for the background lifetime are generated by sampling data  from the mass sidebands. The default fit uses a set of functions to describe the background lifetime. 
\item Pseudo-experiments are  performed  using  two methods of generating the background angles.  The default method uses a set of functions describing the background angles of data without taking correlations between the angles into account.  In the alternative fit 
the background angles are generated using a three dimensional histogram of the sideband-data angles.   
\end{itemize}

\item  {\bf \boldmath{$\Bd$} contribution:} Contamination from $\Bd \ra J/\psi K^{*0}$ and $\Bd \ra J/\psi K\pi$  events mis-reconstructed as \Bst\ are accounted for in the default fit; the fractions of these contributions are fixed to values estimated from selection efficiencies in MC simulation and  decay probabilities from ref.~\cite{PDG}. To estimate the systematic uncertainty arising from the precision of the fraction estimates, the data are fitted with these fractions increased and decreased by $1\sigma$.  The largest shift in the fitted values from the default case is taken as the systematic uncertainty for each parameter of interest.
\end{itemize}
The systematic uncertainties are summarised in table \ref{tab:syst_totals}. In general, pseudo-experiments generated with the default model produce pull-distributions that show a negligible bias, and confirm that the uncertainties are correctly estimated by the fit. The largest average deviation in a residual divided by its fit uncertainty (or ÒpullÓ) is 0.32; the second largest is 0.26, while the remainder where much smaller. ÊThese two largest deviations were added in quadrature to those obtained by varying the model assumptions, resulting for each variable in a total systematic uncertainty shown in table  \ref{tab:syst_totals}.

\section{Results\label{sec:results}}
The full   maximum likelihood fit contains 26 free parameters.  This includes the eight physics parameters: \DGs, \phis, \Gs, $|A_{0}(0)|^2$, $|A_{\parallel}(0)|^2$, $\delta_{||}$, $|A_{S}(0)|^2$ and $\delta_{S}$, and strong phase $\delta_{\perp}$ constrained by external data. The other free parameters in the likelihood function are the \Bs\ signal fraction $f_\textrm{s}$, the parameters describing the $J/\psi \phi$ mass distribution, the parameters describing the decay time and the  angular distributions of the background, the parameters used to describe the estimated decay time uncertainty distributions for signal and background events, and the scale factors between the estimated decay-time and mass uncertainties and their true uncertainties, see equation \ref{eqn:timeErr}.

As discussed in section~\ref{sec:sigPDF}, the strong phase $\delta_{\perp}$ is constrained to the value measured in ref.~\cite{LHCb:2011aa}, as the fit in the absence of flavour tagging is not sufficiently sensitive to this value. The second strong phase, $\delta_{||}$, is fitted very close to its symmetry point at $\pi$. Pull studies, based on pseudo-experiments using  input values determined  from the fit to  data,  return a non-Gaussian pull distribution for this parameter. 
 For this reason the result for the strong phase  $\delta_{||}$  is  given in the form of a $1\sigma$ confidence interval $[3.04, 3.24]$~rad. The strong phase of the $S$-wave component is fitted relative to $\delta_{\perp}$,  as $\delta_{\perp} - \delta_S = (0.03 \pm 0.13)$~rad.

The number of signal \Bs\ meson candidates extracted from the fit is \nsig. The results and correlations for the measured physics parameters of the   unbinned maximum likelihood fit are given in tables~\ref{tab:FitResults} and~\ref{tab:FitResultsCor}.  Fit projections of the mass, proper decay time and angles are given in figures~\ref{fig:simulFitmass},~\ref{fig:simulFittime} and \ref{fig:AngularProjections} respectively.

\begin{table}[htdp]
\begin{center}
\begin{tabular}{|c|c|c|c|}
\hline
Parameter & Value & Statistical & Systematic \\
 & & uncertainty & uncertainty \\
\hline\hline
 \phis (rad) & \pSfit & \pSstat & \pSsyst \\
 \DGs (ps$^{-1}$) & \DGfit & \DGstat & \DGsyst \\
 \Gs (ps$^{-1}$) & \GSfit & \GSstat & \GSsyst \\
   $|A_{0}(0)|^2$ & \Azesqfit  & \Azesqstat & \Azesqsyst \\
  $|A_{\parallel}(0)|^2$ & \Apasqfit & \Apasqstat & \Apasqsyst \\
$|A_{S}(0)|^2$ & \ASsqfit & \ASsqstat & \ASsqsyst \\
\hline
\end{tabular}
\caption{Fitted values for the physics parameters along with their statistical and systematic uncertainties. \label{tab:FitResults}}
\end{center}
\end{table}

\begin{table}[htdp]
\begin{center}

\begin{tabular}{|c||c|c|c|c|c|c|}
\hline
& \phis    & \DGs & \Gs & $|A_{0}(0)|^2$ & $|A_{\parallel}(0)|^2$ & $|A_{S}(0)|^2$ \\
\hline \hline
\phis  &  1.00 &  $\pDG\;\;\,$     & $\pG$ & $\pAz\;\;\,$ & $\pApa\;\;\,$ & $\pAs$    \\
\DGs  &  & 1.00 & $\DGG\;\;\,$ & $\DGAz$ & $\DGApa$ & $\DGAs$ \\
\Gs & & & 1.00 & $\GAz\;\;\,$ & $\GApa\;\;\,$ & $\GAs$ \\
$|A_{0}(0)|^2$ & & & & 1.00 & $\AzApa\;\;\,$ &  $\AzAs$ \\
$|A_{\parallel}(0)|^2$   & & & & & 1.00 & $\ApaAs$   \\
$|A_{S}(0)|^2$  & & & & & & 1.00\\
\hline
\end{tabular}

\caption{ Correlations between the physics parameters. \label{tab:FitResultsCor}}
\end{center}
\end{table}%
\begin{table}
\begin{center}
\begin{tabular}{l | ccccccc} 
 \hline \hline 
Systematic Uncertainty& \phis (rad) & \DGs (ps$^{-1}$)& \Gs (ps$^{-1}$)  &  $|A_{\parallel}(0)|^2$ & $|A_{0}(0)|^2$ & $ |A_{S}(0)|^2 $ \\  
\hline
Inner Detector alignment              & 0.04  & $<0.001$ & 0.001 & $<0.001$ & $<0.001$ &  $<0.01$  \\
Trigger efficiency & $< 0.01$    & $< 0.001 $    &   0.002   &    $<0.001 $  &    $<0.001 $  &$<0.01$  \\
Default fit model  & $<0.001$ & $ 0.006$ & $<0.001$ & $<0.001$ & $0.001$ & $< 0.01$\\
Signal mass model & 0.02 & 0.002 & $<0.001$ & $< 0.001$ & $< 0.001$  & $<0.01$ \\ 
Background mass model & 0.03 & 0.001 & $<0.001$ &  $0.001 $ & $<0.001$ & $<0.01$\\ 
Resolution model & 0.05 & $< 0.001$ & 0.001 & $<0.001$ & $< 0.001$ & $<0.01$ \\ 
Background lifetime model & 0.02 & 0.002 & $<0.001$ & $<0.001$ & $<0.001$ & $<0.01$ \\ 
Background angles model & 0.05 & 0.007 & 0.003 & 0.007 & 0.008 & 0.02 \\
$\Bd$ contribution    & 0.05 & $ <0.001$ & $<0.001$ & $< 0.001$ & 0.005 & $< 0.01$\\
\hline					  
\bf Total  & \pSsyst & \DGsyst & \GSsyst  & \Apasqsyst & \Azesqsyst  & \ASsqsyst \\
\hline \hline 
\end{tabular}
\end{center}
\caption{Summary of systematic uncertainties assigned to parameters of interest.}
\label{tab:syst_totals}
\end{table}

\begin{figure}[htbp]
\begin{center}
\includegraphics[width=0.65\textwidth]{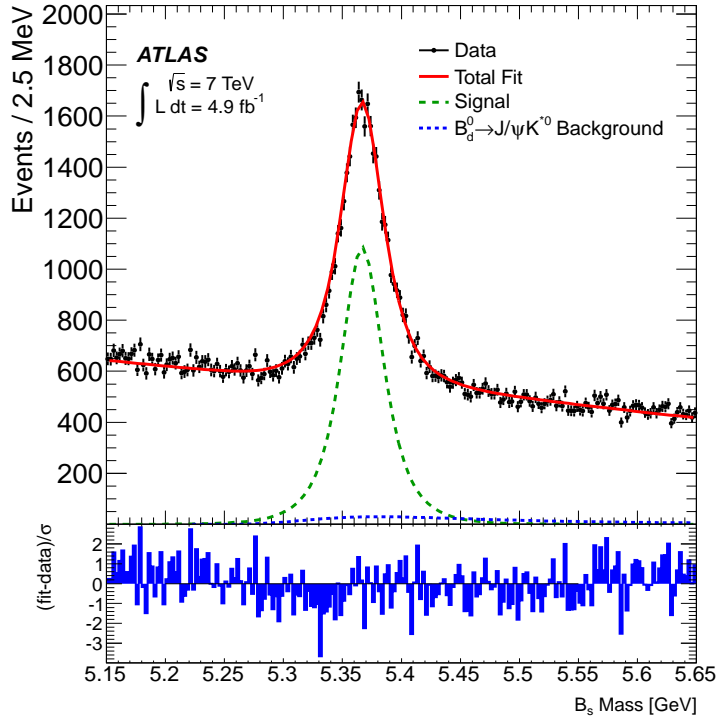}
\caption{Mass fit projection for the \Bs. The pull distribution at the bottom shows the difference between the data and fit value normalised to the data uncertainty.}
\label{fig:simulFitmass}
\end{center}
\end{figure}
\begin{figure}[htbp]
\begin{center}
\includegraphics[width=0.65\textwidth]{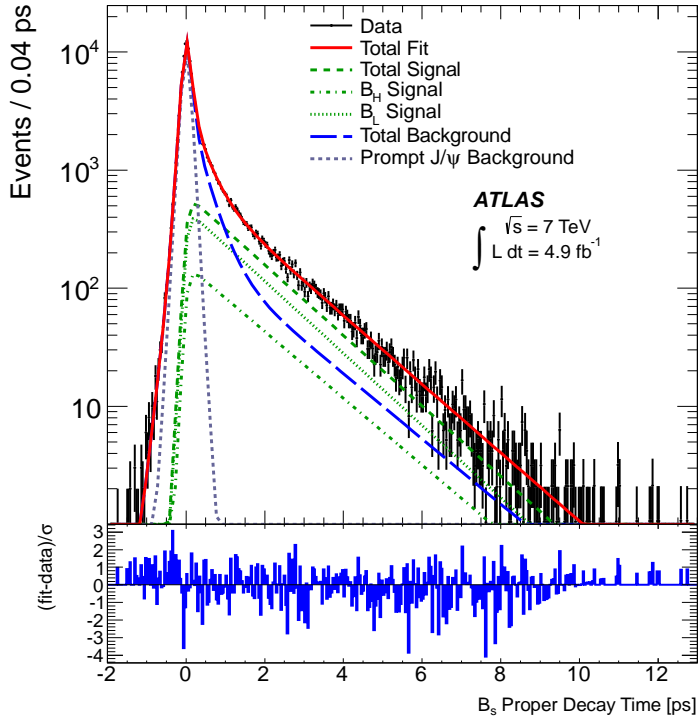}
\caption{ Proper decay time fit projection for the \Bs. The pull distribution at the bottom shows the difference between the data and fit value normalised to the data uncertainty.}
\label{fig:simulFittime}
\end{center}
\end{figure}
\begin{figure}[htbp]
\begin{center}
\includegraphics[width=7.5cm]{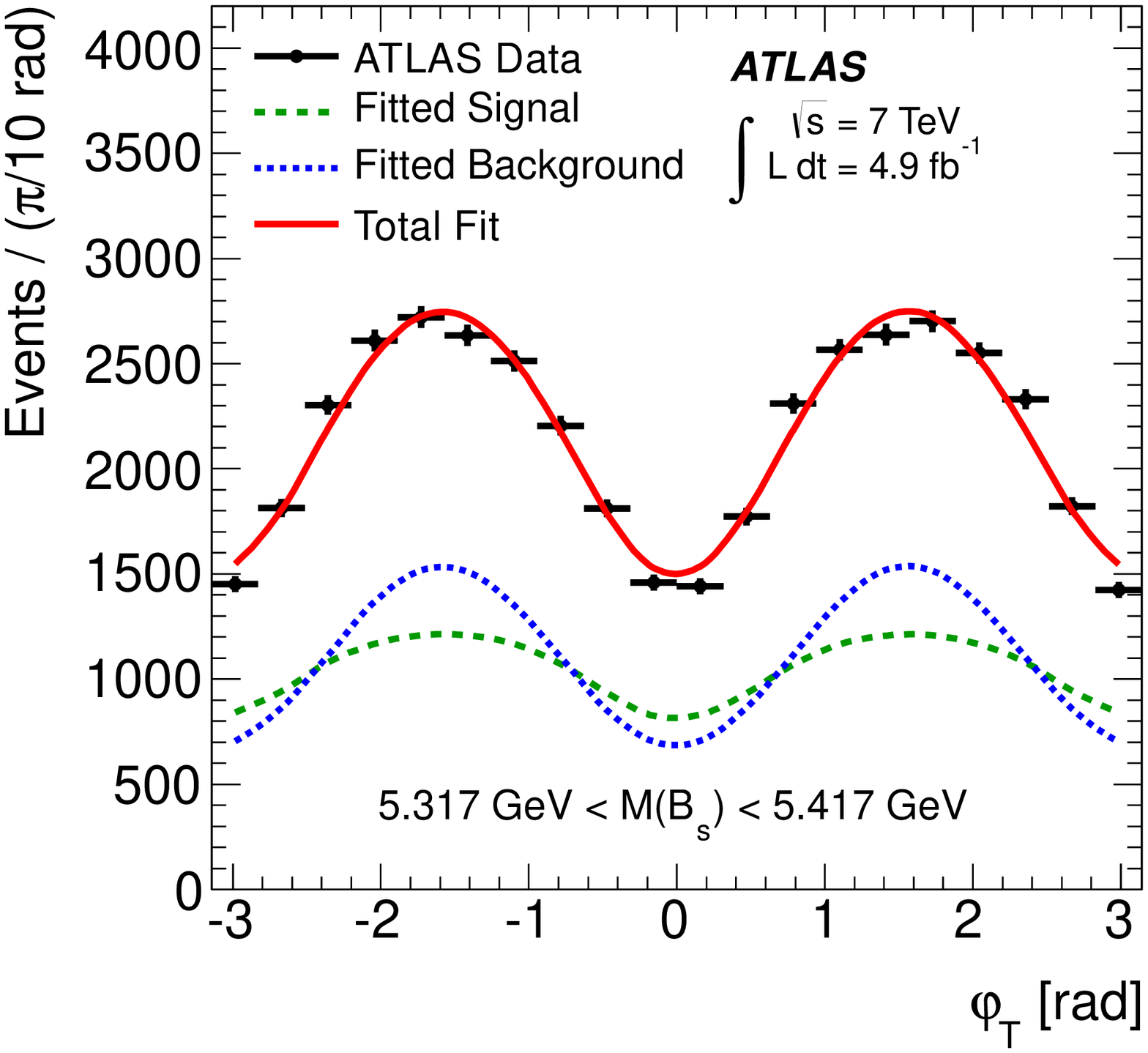}
\includegraphics[width=7.5cm]{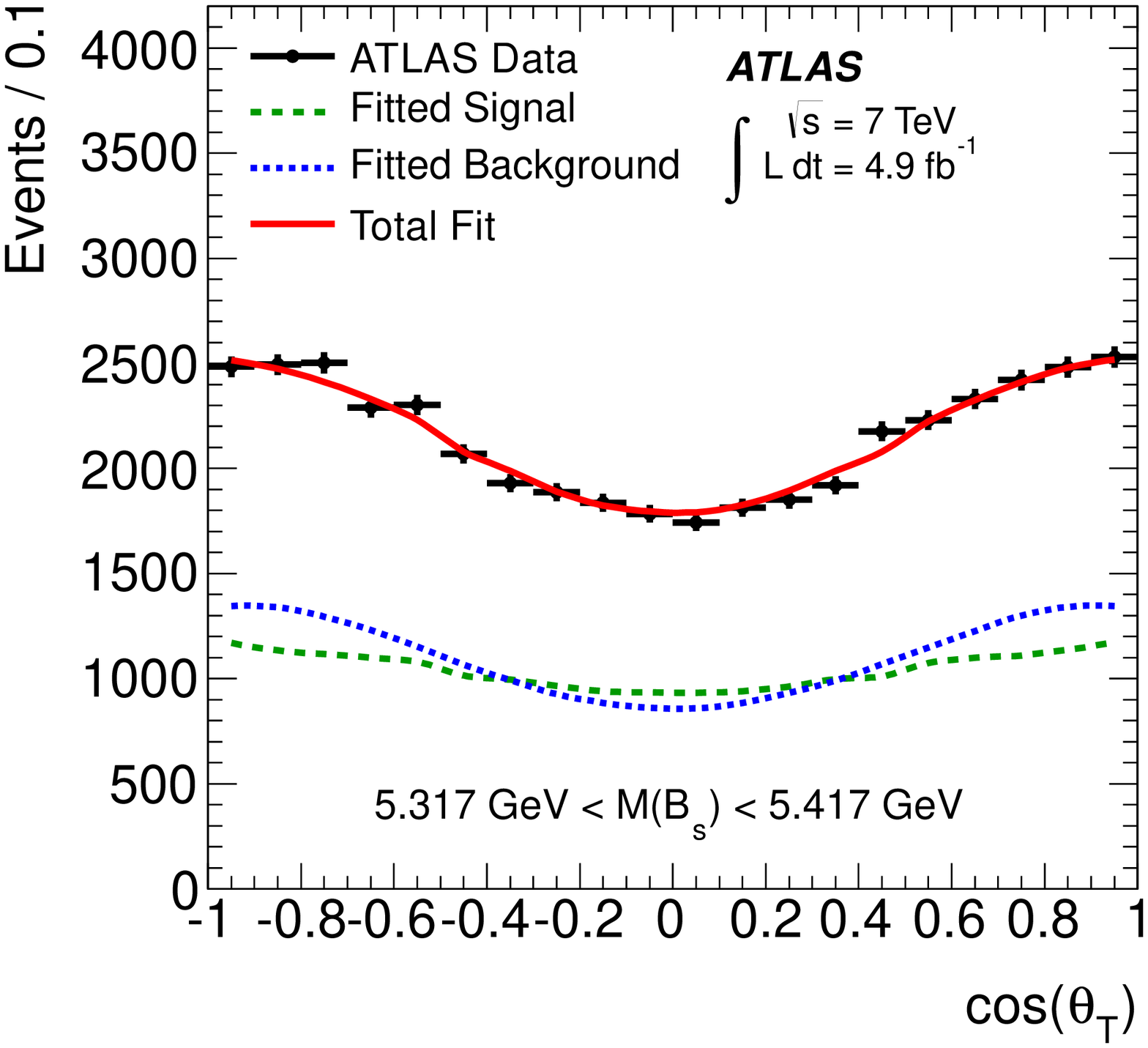}
\includegraphics[width=7.5cm]{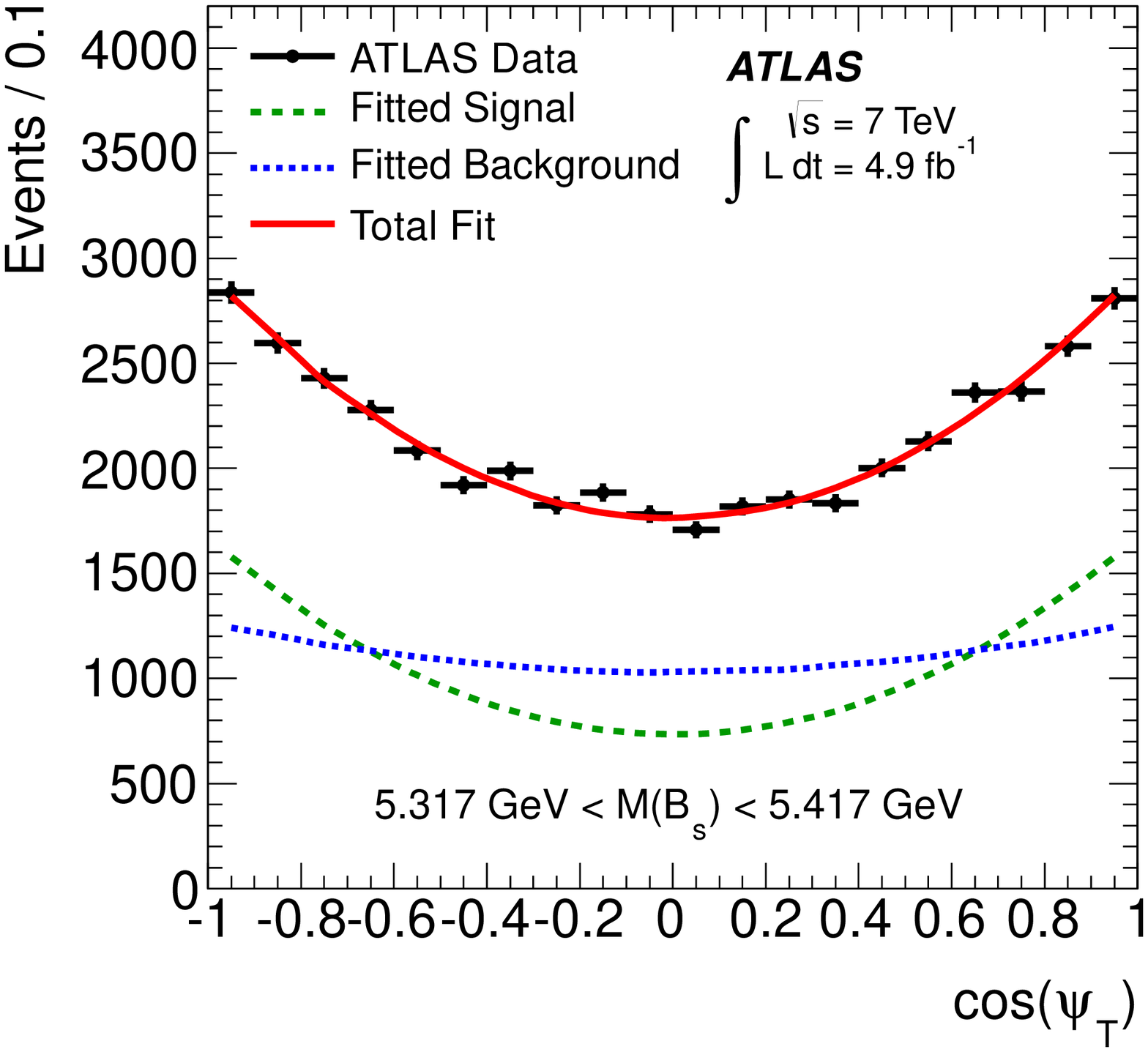}
\caption{Fit projections for transversity angles. (Left): $\varphi_T$, (Right): $\cos\theta_T$, (Bottom): $\cos\psi_T$ for the events with \Bs\ mass 
from signal region (5.317 - 5.417)~GeV. }
\label{fig:AngularProjections}
\end{center}
\end{figure}

\section{Symmetries of the likelihood function and two-dimensional likelihood contours \label{sec:discussion}}
The PDF describing the \Bst\ decay is invariant under the following simultaneous transformations: 
\begin{equation}
  \{\phis, \DGs, \delta_{\perp}, \delta_{\parallel}, \delta_S\}  \rightarrow \{\pi -\phis, -\DGs, \pi - \delta_{\perp}, -\delta_{\parallel}, -\delta_S\}. \nonumber
\end{equation}
In the absence of initial state flavour tagging the PDF is also invariant under  
\begin{equation}
  \{\phis, \DGs, \delta_{\perp}, \delta_{\parallel}, \delta_S\}  \rightarrow \{    -\phis,  \DGs, \pi - \delta_{\perp}, -\delta_{\parallel}, -\delta_S\}
\label{eqn:sym2}
\end{equation}
leading to a fourfold ambiguity.  

The two-dimensional likelihood contours in the \phis\ $-$ \DGs\ plane are calculated allowing all parameters to vary within their physical ranges.  As discussed in section~\ref{sec:results},  the value for the Gaussian constraint on $\delta_{\perp}$ is taken from the LHCb measurement~\cite{LHCb:2011aa}. That paper quotes only two solutions with a positive \phis\ and two \DGs\ values symmetric around zero, by using initial state flavour tagging to eliminate the symmetry defined in equation~\ref{eqn:sym2}. Due to the accurate local determination of \phis\ and \DGs\ in both this measurement and in the LHCb measurement~\cite{LHCb:2011aa}, the other two solutions seen in the ATLAS analysis are not compatible with the observations of the two experiments. As such, two of the four minima fitted in the present non-flavour tagged analysis are excluded from the results presented here. Additionally a solution with negative \DGs\ is excluded following the LHCb measurement~\cite{Aaij:2012eq} which determines the \DGs\ to be positive. Therefore, the  two-dimensional contour plot for \phis\ and  \DGs\ has been computed only for the solution consistent with the previous measurements.  The resulting contours for the 68\%, 90\% and 95\% confidence intervals are produced using a profile likelihood method and are shown in figure \ref{fig:Contour}. 

The systematic errors  are not included in figure \ref{fig:Contour}  but as seen from table \ref{tab:FitResults} they are small  compared to the  statistical errors. The confidence levels are obtained using the corresponding $\Delta\ln{\cal L}$ intervals. Pseudo-experiments are used to study the coverage of the likelihood contours. This test suggests that the statistical uncertainty of our result is overestimated by about $5\%$. No correction to compensate for this overestimation is applied.

\begin{figure}[h]
\begin{center}
\includegraphics[width=0.65\textwidth]{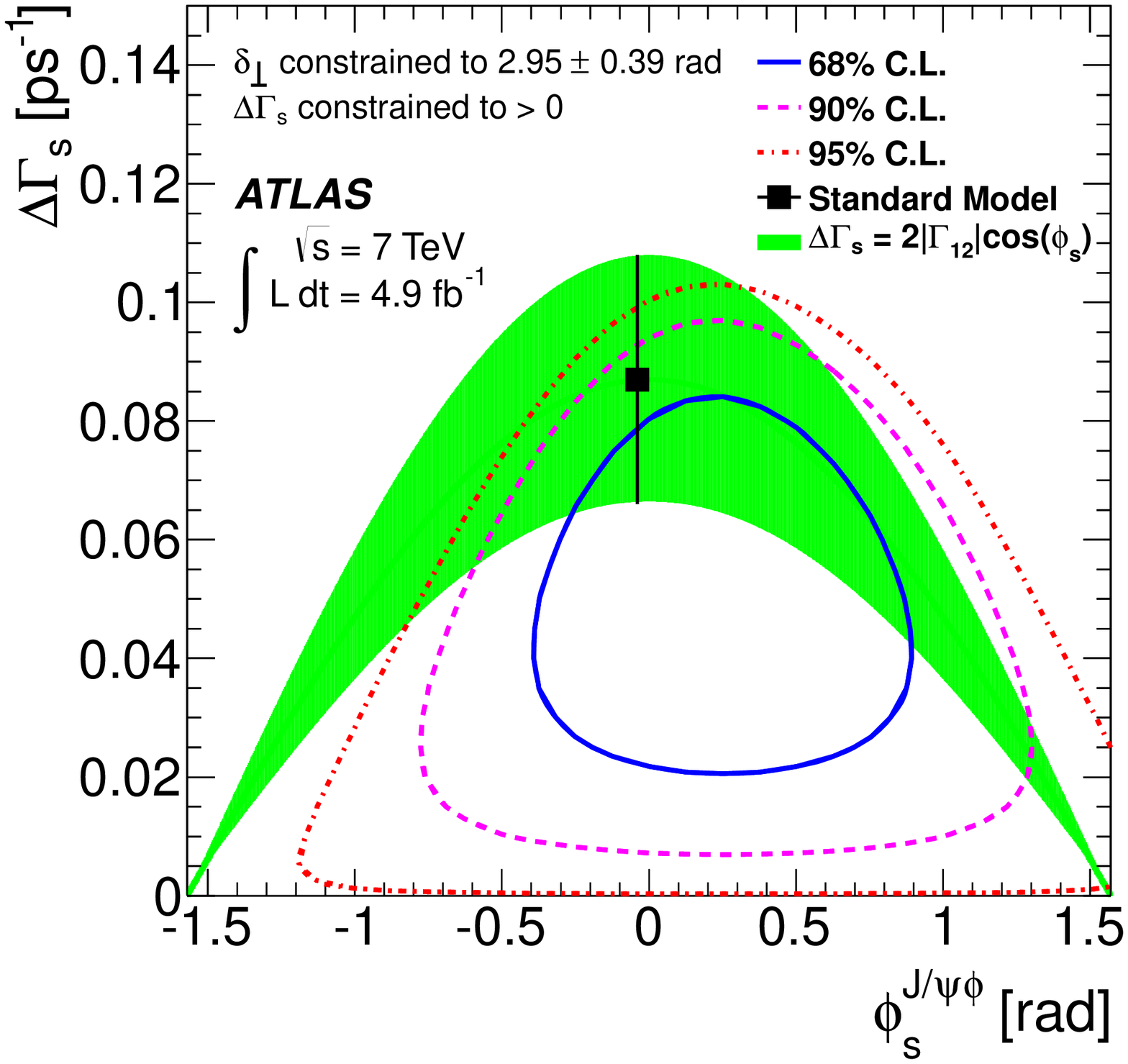}
\caption{Likelihood contours in the \phis\ $-$ \DGs\ plane. 	Three contours show the 68\%, 
90\% and 95\%   confidence intervals (statistical errors only). The green band is the theoretical prediction of mixing- induced $CP$ violation.
The PDF  contains a fourfold ambiguity. Three minima are excluded by applying the constraints from the LHCb measurements~\cite{LHCb:2011aa,Aaij:2012eq}.}
\label{fig:Contour}
\end{center}
\end{figure}
\section{Conclusion}

A measurement of $CP$ violation in \Bsto\ decays  from a \ilumi\ data sample of $pp$ collisions collected with the ATLAS detector during the 2011 $\rts = 7\TeV$ run was presented.  Several parameters describing the \Bs\ meson system are measured. These include the mean \Bs\ lifetime, the decay width difference \DGs\ between the heavy and light mass eigenstates, the transversity amplitudes $|A_{0}(0)|$ and $|A_{\parallel}(0)|$ and the $CP$-violating week phase \phis.  They are consistent with the world average values.

The measured values, for the minimum resulting from $\delta_{\perp}$ constrained to the LHCb value of $2.95 \pm 0.39$ rad~\cite{LHCb:2011aa} and \DGs\ being constrained to be positive following LHCb measurement~\cite{Aaij:2012eq}, are:
\begin{alignat*}{3}
\phis & = & \pSfit \,~  & \ \pm & \pSstat \,~~ \rm{(stat.)} & \pm  \ \pSsyst \,~~ \rm{(syst.)}~rad  \nonumber \\
\DGs &  = & ~ \DGfit  & \ \pm  & ~ \DGstat~\rm{(stat.)}  & \pm   ~  \DGsyst~\rm{(syst.)~ps}^{-1} \nonumber \\
\Gs & = & \GSfit & \ \pm & \GSstat ~\rm{(stat.)} & \pm \ \GSsyst ~\rm{(syst.)~ps}^{-1} \nonumber \\
|A_{0}(0)|^2 & = & \Azesqfit & \ \pm & \Azesqstat ~\rm{(stat.)}& \pm \ \Azesqsyst~\rm{(syst.)} \nonumber \\
|A_{\parallel}(0)|^2 & = & \Apasqfit & \ \pm & \Apasqstat ~\rm{(stat.)}&  \pm \ \Apasqsyst~\rm{(syst.)} \nonumber \\
\end{alignat*}
These values are consistent with theoretical expectations, in particular \phis\ is within $1\sigma$ of the expected value in the Standard Model.
A likelihood contour in the \phis\ $-$ \DGs\ plane is also provided for the minimum compatible with the LHCb measurements~\cite{LHCb:2011aa,Aaij:2012eq}.
The fraction of $S$-wave $KK$ or $f_{0}$ contamination is measured to be consistent with zero, at $|A_S(0)|^2 = \ASsqfit \pm \ASsqstat$.

\input{Acknowledgements}

\bibliographystyle{JHEP}
\bibliography{atlaspaper}
\clearpage
\input{atlas_authlist} 
\end{document}

%% file: keys.tex
\newcommand{\ilumi} {$ 4.9\, \rm fb^{-1} $}

\newcommand{\DGfit}     { 0.053 }

\newcommand{\DGstat}    { 0.021 }
\newcommand{\DGsyst}    { 0.010 }

\newcommand{\pSfit}     { 0.22 }

\newcommand{\pSstat}    { 0.41 }
\newcommand{\pSsyst}    { 0.10 }

\newcommand{\GSfit}     { 0.677 }

\newcommand{\GSstat}    { 0.007 }
\newcommand{\GSsyst}    { 0.004 }

\newcommand{\Azesqfit}     { 0.528 }

\newcommand{\Azesqstat}    { 0.006 }
\newcommand{\Azesqsyst}    { 0.009 }

\newcommand{\Apasqfit}     { 0.220 }

\newcommand{\Apasqstat}    { 0.008 }
\newcommand{\Apasqsyst}    { 0.007 }

\newcommand{\ASsqfit}      { 0.02 }

\newcommand{\ASsqstat}     { 0.02 }
\newcommand{\ASsqsyst}     { 0.02 }

\newcommand{\nsig} {\ensuremath{ 22690 \pm 160 }}

\newcommand{\pDG}   { -0.13 }
\newcommand{\pG}    {  0.38 }
\newcommand{\pAz}   { -0.03 }
\newcommand{\pApa}  { -0.04 }
\newcommand{\pAs}   {  0.02 }

\newcommand{\DGG}   { -0.60 }
\newcommand{\DGAz}  {  0.12 }
\newcommand{\DGApa} {  0.11 }
\newcommand{\DGAs}  {  0.10 }

\newcommand{\GAz}   { -0.06 }
\newcommand{\GApa}  { -0.10 }
\newcommand{\GAs}   {  0.04 }

\newcommand{\AzApa} { -0.30 }
\newcommand{\AzAs}  {  0.35 }

\newcommand{\ApaAs} {  0.09 }

\newcommand{\cutMassJpsiBB} {$ (2.959-3.229) ~\rm GeV $}
\newcommand{\cutMassJpsiEB} {$ (2.913-3.273) ~\rm GeV $}
\newcommand{\cutMassJpsiEE} {$ (2.852-3.332) ~\rm GeV $}

\newcommand{\cutChiJpsi}   {\chisq $< 10$}

\newcommand{\cutMassPhi} {$ 1.0085 \; {\rm GeV} < m(K^+ K^-) < 1.0305 \; {\rm GeV} $}
\newcommand{\cutChiBs}   {\chisq $< 3$}

\newcommand{\BspdgNoerr}  {$ 5.3663 ~\rm GeV $}

\newcommand{\Bst}{\ensuremath{ \Bs \ra \JpsiPhi }} 
\newcommand{\Bstbold}{\boldmath{ \Bs \ra \JpsiPhi }} 
 
\newcommand{\Bsto}{\ensuremath{ \Bs \ra J/\psi (\mumu) \phi(K^+K^-) }} 
 
\newcommand{\Bdt}{\ensuremath{ \Bd \ra J/\psi K^{*} }} 

\newcommand{\BstoKK}{\ensuremath{ \Bs \ra J/\psi K^+K^- (f_{0}) }} 
\newcommand{\JpsiPhi}{\ensuremath{ J/\psi\phi }}

\newcommand{\Gs}{\ensuremath{ \Gamma_{s}     }}
\newcommand{\DGs}{\ensuremath{ \Delta\Gamma_{s}     }} 

\newcommand{\phis}{\ensuremath{\phi_{s}  }} 
\newcommand{\dms}{\ensuremath{\Delta m_{s}  }} 
\newcommand{\BH}{\ensuremath{B_{H}  }} 
\newcommand{\BL}{\ensuremath{B_{L}  }} 
\newcommand{\GL}{\ensuremath{\Gamma_{L}  }}

\def \ra{\ensuremath{\rightarrow}}

\newcommand{\BsaBs}{\ensuremath{\Bs - \overline{\Bs}}\space} 
\newcommand{\B}{\ensuremath{ B}\space  }

\newcommand{\aBsn}{\ensuremath{\overline{\Bs}}}

\newcommand{\Jpsipa}{\Jpsi \ }

\newcommand{\psubt}{ \pt \space}

\newcommand{\chisq}{$\rm \chi^{2}$/d.o.f. }

\newcommand{\Jpsitomm}{\ensuremath{ \Jpsi \ra \mumu }\space}

\newcommand{\phiKK}{\ensuremath{ \phi \ra \kplus \kminus}\space}



%% file: Acknowledgements.tex



\section{Acknowledgements}

We thank CERN for the very successful operation of the LHC, as well as the
support staff from our institutions without whom ATLAS could not be
operated efficiently.

We acknowledge the support of ANPCyT, Argentina; YerPhI, Armenia; ARC,
Australia; BMWF and FWF, Austria; ANAS, Azerbaijan; SSTC, Belarus; CNPq and FAPESP,
Brazil; NSERC, NRC and CFI, Canada; CERN; CONICYT, Chile; CAS, MOST and NSFC,
China; COLCIENCIAS, Colombia; MSMT CR, MPO CR and VSC CR, Czech Republic;
DNRF, DNSRC and Lundbeck Foundation, Denmark; EPLANET, ERC and NSRF, European Union;
IN2P3-CNRS, CEA-DSM/IRFU, France; GNSF, Georgia; BMBF, DFG, HGF, MPG and AvH
Foundation, Germany; GSRT and NSRF, Greece; ISF, MINERVA, GIF, DIP and Benoziyo Center,
Israel; INFN, Italy; MEXT and JSPS, Japan; CNRST, Morocco; FOM and NWO,
Netherlands; BRF and RCN, Norway; MNiSW, Poland; GRICES and FCT, Portugal; MERYS
(MECTS), Romania; MES of Russia and ROSATOM, Russian Federation; JINR; MSTD,
Serbia; MSSR, Slovakia; ARRS and MVZT, Slovenia; DST/NRF, South Africa;
MICINN, Spain; SRC and Wallenberg Foundation, Sweden; SER, SNSF and Cantons of
Bern and Geneva, Switzerland; NSC, Taiwan; TAEK, Turkey; STFC, the Royal
Society and Leverhulme Trust, United Kingdom; DOE and NSF, United States of
America.

The crucial computing support from all WLCG partners is acknowledged
gratefully, in particular from CERN and the ATLAS Tier-1 facilities at
TRIUMF (Canada), NDGF (Denmark, Norway, Sweden), CC-IN2P3 (France),
KIT/GridKA (Germany), INFN-CNAF (Italy), NL-T1 (Netherlands), PIC (Spain),
ASGC (Taiwan), RAL (UK) and BNL (USA) and in the Tier-2 facilities
worldwide.

%% file: atlas_authlist.tex
\begin{flushleft}
{\Large The ATLAS Collaboration}

\bigskip

G.~Aad$^{\rm 48}$,
T.~Abajyan$^{\rm 21}$,
B.~Abbott$^{\rm 111}$,
J.~Abdallah$^{\rm 12}$,
S.~Abdel~Khalek$^{\rm 115}$,
A.A.~Abdelalim$^{\rm 49}$,
O.~Abdinov$^{\rm 11}$,
R.~Aben$^{\rm 105}$,
B.~Abi$^{\rm 112}$,
M.~Abolins$^{\rm 88}$
% ...
\bigskip
\\
% ...
$^{11}$ Institute of Physics, Azerbaijan Academy of Sciences, Baku, Azerbaijan\\
$^{12}$ Institut de F{\'\i}sica d'Altes Energies and Departament de F{\'\i}sica de la Universitat Aut{\`o}noma de Barcelona and ICREA, Barcelona, Spain\\
% ...
$^{21}$ Physikalisches Institut, University of Bonn, Bonn, Germany\\
% ...
$^{48}$ Fakult{\"a}t f{\"u}r Mathematik und Physik, Albert-Ludwigs-Universit{\"a}t, Freiburg, Germany\\
$^{49}$ Section de Physique, Universit{\'e} de Gen{\`e}ve, Geneva, Switzerland\\
% ...
$^{88}$ Department of Physics and Astronomy, Michigan State University, East Lansing MI, United States of America\\
% ...
$^{105}$ Nikhef National Institute for Subatomic Physics and University of Amsterdam, Amsterdam, Netherlands\\
% ...
$^{111}$ Homer L. Dodge Department of Physics and Astronomy, University of Oklahoma, Norman OK, United States of America\\
$^{112}$ Department of Physics, Oklahoma State University, Stillwater OK, United States of America\\
% ...
$^{115}$ LAL, Universit{\'e} Paris-Sud and CNRS/IN2P3, Orsay, France\\
% ...
\end{flushleft}
